\documentclass[12pt]{article}
\usepackage{makeidx}
\usepackage{amsmath}
\usepackage{amsfonts}
\usepackage{amssymb}
\usepackage{epsfig}

\newcounter{defin}
\newcounter{lemma}
\newcounter{theorem}
\newcounter{corollary}
\newcounter{proposition}
\newcounter{example}

\begin{document}

\title{On estimates of the Bures distance between bosonic Gaussian states%
}
\author{Holevo~A.\,S. \\
Steklov Mathematical Institute, \\
Russian Academy of Sciences, Moscow, Russia}
\date{}
\maketitle

\begin{abstract}
The aim of the present note is to show that the method of our paper \textsf{%
ArXiv:2408.11400} with minor extra efforts can be extended to obtain upper
bounds for the Bures distance between quantum Gaussian states. We argue that
these bounds are better adapted to the Bures distance and hence to the state
estimation with the Bures distance rather than that with the trace-norm
distance.

\end{abstract}

\section{Introduction}

In the paper \cite{lami} estimates for the trace-norm distance $\left\Vert
\rho _{1}-\rho _{2}\right\Vert _{1}$ between two quantum Gaussian states in
terms of their parameters -- the mean vectors and covariance matrices were
derived and used to evaluate the sample complexity of learning
energy-constrained Gaussian states. In our posting \cite{ho} we suggested
alternative estimates with a more direct proof based on the quantity $%
\mathrm{Tr}\sqrt{\rho _{1}}\sqrt{\rho _{2}}$ (which we called quantum states
overlap) and the corresponding inequalities for of the trace-norm distance
from \cite{tmf}. More recently, the sequel \cite{lami2} of \cite{lami}
appeared where the estimates of \cite{lami} were substantially improved by
using still a different ``gradient'' method.

The choice of the trace-norm distance was substantiated in \cite{lami} by
its relevance to the optimal discrimination of quantum states. Meanwhile
other estimates based on the quantum relative entropy and Pinsker inequality
targeted to Hamiltonian learning were demonstrated in \cite{rouze}. Another
important distinguishability measure is the Bures distance $d_{B}\left( \rho
_{1},\rho _{2}\right) $ defining Riemannian metric associated with the left
logarithmic derivative of states and the quantum Fisher information, and as
such also relevant to the state estimation and learning, cf. \cite{wilde}.

The aim of the present note is to show that the method of our paper \cite{ho}
with minor extra efforts can be extended to obtain similar upper bounds for
the Bures distance. Moreover, we argue that these bounds are better adapted
to the Bures distance and hence to the state estimation with the Bures
distance rather than that with the trace-norm distance.

Throughout this note $\Vert \cdot \Vert _{1}$ denotes \textit{the trace norm}
of a matrix or an operator on a Hilbert space, $\Vert \cdot \Vert _{2}$ --
either \textit{the Hilbert-Schmidt norm} of an operator or \textit{Euclidean
norm} of a real vector, and $\Vert \cdot \Vert $ --\textit{\ the operator
norm.} Otherwise we use definitions and notations from our paper \cite{ho}
without detailed explanation.

\section{Distances and fidelities}


Our approach in \cite{ho} uses the inequalities%
\begin{equation}
2\left( 1-\mathrm{Tr}\sqrt{\rho _{1}}\sqrt{\rho _{2}}\right) \leq \left\Vert
\rho _{1}-\rho _{2}\right\Vert _{1}\leq 2\sqrt{1-\left( \mathrm{Tr}\sqrt{%
\rho _{1}}\sqrt{\rho _{2}}\right) ^{2}},  \label{basic}
\end{equation}%
where $\rho _{1},\rho _{2}$ are density operators. The proof given in \cite%
{tmf} is partly based on estimates from \cite{powers}. (see e.g. Example
2.25 in \cite{h}). The inequalities (\ref{basic}) are useful but less known
than the Fuchs-de Graaf inequalities \cite{fuchs}%
\begin{equation}
2\left( 1-\mathrm{Tr}\left\vert \sqrt{\rho _{1}}\sqrt{\rho _{2}}\right\vert
\right) \leq \left\Vert \rho _{1}-\rho _{2}\right\Vert _{1}\leq 2\sqrt{%
1-\left( \mathrm{Tr}\left\vert \sqrt{\rho _{1}}\sqrt{\rho _{2}}\right\vert
\right) ^{2}},  \label{fdg}
\end{equation}%
involving the fidelity $\left( \mathrm{Tr}\left\vert \sqrt{\rho _{1}}\sqrt{%
\rho _{2}}\right\vert \right) ^{2}$ between the quantum states $\rho
_{1},\rho _{2}$. Note that the first inequality is stronger in (\ref{basic})
\ while the second -- in (\ref{fdg}), because
\begin{equation}
\mathrm{Tr}\sqrt{\rho _{1}}\sqrt{\rho _{2}}\leq \mathrm{Tr}\left\vert \sqrt{%
\rho _{1}}\sqrt{\rho _{2}}\right\vert .  \label{abs}
\end{equation}

An important distinguishability measure for quantum states is the Bures
distance (cf. sec. 10.2 of \cite{h})
\begin{equation*}
d_{B}\left( \rho _{1},\rho _{2}\right) =\sqrt{2\left( 1-\mathrm{Tr}%
\left\vert \sqrt{\rho _{1}}\sqrt{\rho _{2}}\right\vert \right) }.
\end{equation*}%
%
%
We will rely upon the upper bound in terms of the states overlap following
from (\ref{abs})
\begin{equation}
d_{B}\left( \rho _{1},\rho _{2}\right) \leq \sqrt{2\left( 1-\mathrm{Tr}\sqrt{%
\rho _{1}}\sqrt{\rho _{2}}\right) }.  \label{bur}
\end{equation}%
\ Note that (\ref{fdg})\ implies
\begin{equation}
d_{B}\left( \rho _{1},\rho _{2}\right) ^{2}\leq \left\Vert \rho _{1}-\rho
_{2}\right\Vert _{1}\leq 2d_{B}\left( \rho _{1},\rho _{2}\right) .
\label{tnb}
\end{equation}

In the case of pure states $\rho _{j}=|\psi _{j}\rangle \langle \psi
_{j}|;j=1,2,$
\begin{equation*}
\left\vert \left\langle \psi _{1}|\psi _{2}\right\rangle \right\vert ^{2}=%
\mathrm{Tr}\rho _{1}\rho _{2}=\mathrm{Tr}\sqrt{\rho _{1}}\sqrt{\rho _{2}}%
=\left( \mathrm{Tr}\left\vert \sqrt{\rho _{1}}\sqrt{\rho _{2}}\right\vert
\right) ^{2}
\end{equation*}%
implying exact expressions (see e.g. Lemma 10.9 in \cite{h})
\begin{eqnarray}
\left\Vert \rho _{1}-\rho _{2}\right\Vert _{1} &=&2\sqrt{1-\mathrm{Tr}\sqrt{%
\rho _{1}}\sqrt{\rho _{2}}},  \label{basic3} \\
d_{B}\left( \rho _{1},\rho _{2}\right) &=&\sqrt{2\left( 1-\sqrt{\mathrm{Tr}%
\sqrt{\rho _{1}}\sqrt{\rho _{2}}}\right) }.  \label{basic5}
\end{eqnarray}

\section{The upper bounds}


We consider bosonic system with $s$ modes and the canonical commutation
matrix
\begin{equation}
\Delta =\mathrm{diag}\left[
\begin{array}{cc}
0 & 1 \\
-1 & 0%
\end{array}%
\right] _{j=1,\dots ,s}.
\end{equation}%
Let $\rho _{m_{1},\alpha }$ and $\rho _{m_{2},\beta }$ be two quantum
Gaussian states with the mean vectors $m_{1},m_{2}$ and the covariance
matrices $\alpha $, $\beta $. Defining the matrix

\begin{equation*}
\hat{\alpha}=\alpha \left( I_{2s}+\sqrt{I_{2s}+\left( 2\Delta ^{-1}\alpha
\right) ^{-2}}\right)
\end{equation*}%
we have (Eq.(29) in \cite{ho}, also \cite{tmf})
\begin{equation}
\mathrm{Tr}\sqrt{\rho _{m_{1,}\alpha }}\sqrt{\rho _{m_{2},\beta }}=\frac{%
\left( \det \hat{\alpha}\det \hat{\beta}\right) ^{1/4}}{\det \sigma (\alpha
,\beta )^{1/2}}\exp \left( -\frac{1}{4}m^{t}\sigma (\hat{\alpha},\hat{\beta}%
)^{-1}m\right) ,  \label{overl}
\end{equation}%
where%
\begin{equation*}
\sigma (\alpha ,\beta )=\frac{\hat{\alpha}+\hat{\beta}}{2},\quad
m=m_{1}-m_{2}.
\end{equation*}%
Arguing as in \cite{ho} we evaluate the factor before the exponent to obtain
\begin{equation}
1-\mathrm{Tr}\sqrt{\rho _{\alpha }}\sqrt{\rho _{m,\beta }}\leq \frac{1}{4}%
\left[ m^{t}\sigma (\alpha ,\beta )^{-1}m+\mathrm{Tr\,}\delta (\alpha ,\beta
)\right] .  \label{ineq6}
\end{equation}%
where%
\begin{equation*}
\mathrm{Tr\,}\delta (\alpha ,\beta )=\frac{1}{4}\mathrm{Tr}\left( \hat{\alpha%
}-\hat{\beta}\right) \left( \hat{\beta}^{-1}-\hat{\alpha}^{-1}\right) .
\end{equation*}%
Denoting $\Upsilon _{\alpha }=\sqrt{I_{2s}+\left( 2\Delta ^{-1}\alpha
\right) ^{-2}}$ we have as in \cite{ho} Eq. (34-36)
\begin{eqnarray}
&&\mathrm{Tr\,}\delta (\alpha ,\beta )=\frac{1}{4}\mathrm{Tr}\left( \hat{%
\alpha}-\hat{\beta}\right) \left( \hat{\beta}^{-1}-\hat{\alpha}^{-1}\right)
\label{ea} \\
&=&\mathrm{Tr}[\left( \alpha -\beta \right) +\left( \alpha \Upsilon _{\alpha
}-\beta \Upsilon _{\beta }\right) ]\Delta ^{-1}[\left( \alpha -\beta \right)
-\left( \alpha \Upsilon _{\alpha }-\beta \Upsilon _{\beta }\right) ]\Delta
^{-1}  \notag \\
&=&\mathrm{Tr}[\left( \alpha -\beta \right) \Delta ^{-1}]^{2}-\mathrm{Tr}%
[\left( \alpha \Upsilon _{\alpha }-\beta \Upsilon _{\beta }\right) \Delta
^{-1}]^{2}  \label{eb} \\
&\leq &\mathrm{Tr}[\alpha -\beta ]^{2}+\mathrm{Tr}[\alpha \Upsilon _{\alpha
}-\beta \Upsilon _{\beta }]^{2}.  \label{ec}
\end{eqnarray}%
For $\sigma \left( \alpha ,\beta \right) $ we have similarly to \cite{ho}
Eq. (38)
\begin{equation}
m^{t}\sigma \left( \alpha ,\beta \right) ^{-1}m\leq 2(\left\Vert \alpha
\right\Vert +\left\Vert \beta \right\Vert )\left\Vert m\right\Vert _{2}^{2},
\label{mean}
\end{equation}

By using the inequalities (\ref{bur}), (\ref{tnb}) and (\ref{ineq6}) we have

\textbf{Proposition.} \ \textit{For two Gaussian states, the distances are
upperbounded as}
\begin{equation}
\left\Vert \rho _{m_{1,}\alpha }-\rho _{_{m_{2},\beta }}\right\Vert _{1}\leq
2d_{B}\left( \rho _{m_{1,}\alpha },\rho _{_{m_{2},\beta }}\right) \leq 2%
\sqrt{\frac{1}{2}\left[ m^{t}\sigma (\alpha ,\beta )^{-1}m+\mathrm{Tr\,}%
\delta (\alpha ,\beta )\right] },  \label{basic2}
\end{equation}%
%

For pure Gaussian states we can use (\ref{basic5}) instead of (\ref{bur}).
Also $\left( 2\Delta ^{-1}\alpha \right) ^{2}=-I_{2s},$ that is $\Upsilon
_{\alpha }=0$ and $\hat{\alpha}=\alpha .$ Similarly $\hat{\beta}=\beta $ and
$\Upsilon _{\beta }=0$. Then by using (\ref{mean}) and (\ref{ec}) we obtain
\begin{eqnarray}
\left\Vert \rho _{m_{1,}\alpha }-\rho _{_{m_{2},\beta }}\right\Vert _{1}
&\leq &2d_{B}\left( \rho _{m_{1},\alpha },\rho _{_{m_{2},\beta }}\right)
\label{E1} \\
&\leq &\sqrt{2\left( \left\Vert \alpha \right\Vert +\left\Vert \beta
\right\Vert \right) \left\Vert m_{1}-m_{2}\right\Vert _{2}^{2}+\left\Vert
\alpha -\beta \right\Vert _{2}^{2}}.  \notag
\end{eqnarray}

For gauge invariant (\textquotedblleft passive\textquotedblright ) states $%
m_{1}=m_{2}=0$, and $\alpha ,$ $\beta $ commute with $\Delta $. Taking into
account that $\Delta ^{2}=-I_{2s}$, the expression (\ref{eb}) turns into%
\begin{eqnarray}
\mathrm{Tr\,}\delta (\alpha ,\beta ) &=&-\mathrm{Tr}[\alpha -\beta ]^{2}+%
\mathrm{Tr}[\alpha \Upsilon _{\alpha }-\beta \Upsilon _{\beta }]^{2}
\label{sqrt} \\
&=&.-\left\Vert \alpha -\beta \right\Vert _{2}^{2}+\left\Vert \sqrt{\alpha
^{2}-\left( 1/4\right) I_{2s}}-\sqrt{\beta ^{2}-\left( 1/4\right) I_{2s}}%
\right\Vert _{2}^{2}.  \notag
\end{eqnarray}%
Arguing as in \cite{ho} Eq. (42-43) to evaluate the second term, we obtain
\begin{equation}
\left\Vert \rho _{\alpha }-\rho _{_{\beta }}\right\Vert _{1}\leq
2d_{B}\left( \rho _{\alpha },\rho _{_{\beta }}\right) \leq \sqrt{2[\left(
\left\Vert \alpha \right\Vert +\left\Vert \beta \right\Vert \right)
\left\Vert \alpha -\beta \right\Vert _{1}-\left\Vert \alpha -\beta
\right\Vert _{2}^{2}]}.  \label{E2}
\end{equation}

For general Gaussian states we can obtain the estimate similar to (3) in
\cite{ho}.

\section{Discussion}

An interesting special case is when $\rho _{_{\beta }}=\left\vert
0\right\rangle \left\langle 0\right\vert $ is the (unique) gauge-invariant
pure Gaussian state, so that $\beta =\left( 1/2\right) I_{2s}$ and $\rho
_{\alpha }$ is Gaussian thermal state with $\alpha $ close to $\beta .$
Equation (\ref{sqrt}) implies
\begin{eqnarray*}
\mathrm{Tr\,}\delta (\alpha ,\beta ) &=&-\mathrm{Tr}[\alpha -\left(
1/2\right) I_{2s}]^{2}+\mathrm{Tr}[\alpha ^{2}-\left( 1/4\right) I_{2s}] \\
&=&\mathrm{Tr}[\alpha -\left( 1/2\right) I_{2s}]=\left\Vert \alpha -\beta
\right\Vert _{1},
\end{eqnarray*}%
which according to (\ref{basic2}) leads to
\begin{equation}
\left\Vert \rho _{\alpha }-\rho _{_{\beta }}\right\Vert _{1}\leq
2d_{B}\left( \rho _{\alpha },\rho _{_{\beta }}\right) \leq \sqrt{2\left\Vert
\alpha -\beta \right\Vert _{1}}.  \label{three}
\end{equation}%
The greatest eigenvalue of $\rho _{\alpha }$ is
\begin{equation*}
\lambda _{0}=\left\langle 0\right\vert \rho _{\alpha }\left\vert
0\right\rangle =\left( \mathrm{Tr}\sqrt{\rho _{\alpha }}\sqrt{\rho _{_{\beta
}}}\right) ^{2}=\left( \mathrm{Tr}\left\vert \sqrt{\rho _{\alpha }}\sqrt{%
\rho _{_{\beta }}}\right\vert \right) ^{2}.
\end{equation*}%
One can see (e.g. from (\ref{overl})) that%
\begin{equation*}
\lambda _{0}=\left[ \det \left( \alpha +\left( 1/2\right) I_{2s}\right) %
\right] ^{-1/2}=\exp \left[ -\frac{1}{2}\mathrm{Tr\ln }\left( I_{2s}+\left(
\alpha -\left( 1/2\right) I_{2s}\right) \right) \right] .
\end{equation*}%
In the limit $\left\Vert \alpha -\beta \right\Vert _{1}\equiv \mathrm{Tr}%
[\alpha -\left( 1/2\right) I_{2s}]\rightarrow 0$ this implies%
\begin{equation*}
\lambda _{0}=1-\frac{1}{2}\left\Vert \alpha -\beta \right\Vert _{1}+o\left(
\left\Vert \alpha -\beta \right\Vert _{1}\right) .
\end{equation*}%
Hence%
\begin{eqnarray*}
\left\Vert \rho _{\alpha }-\rho _{_{\beta }}\right\Vert _{1} &=&2(1-\lambda
_{0})\sim \left\Vert \alpha -\beta \right\Vert _{1}, \\
2d_{B}\left( \rho _{\alpha },\rho _{_{\beta }}\right)  &=&2\sqrt{2(1-\sqrt{%
\lambda _{0}})}\sim \sqrt{2\left\Vert \alpha -\beta \right\Vert _{1}}
\end{eqnarray*}%
Looking at (\ref{three}) one sees that this method allows to obtain the
upper bound wnich is tight in the limit $\left\Vert \alpha -\beta
\right\Vert _{1}\rightarrow 0$ with respect to the Bures distance, but not
for the trace-norm distance. The upper bound for the trace-norm distance
which scales as $\left\Vert \alpha -\beta \right\Vert _{1}$ was demonstrated
in the paper \cite{lami2} improving the bound of \cite{lami}. On the other
hand, this also implies that the improved sample complexity estimate from
\cite{lami2} is not valid for the state learning with the Bures distance,
for which one should use rather estimates similar to the initial ones
obtained in \cite{lami}.



\end{document}